\documentclass[letterpaper, 10 pt, conference]{ieeeconf}  

\IEEEoverridecommandlockouts                              

\usepackage{graphics} 
\usepackage{epsfig} 
\usepackage{times} 
\usepackage{amsmath} 
\usepackage{amssymb}  
\usepackage{fancyhdr}
\usepackage[dvipsnames]{xcolor}

\usepackage{accents}
\newlength{\dhatheight}

\definecolor{antiquefuchsia}{rgb}{0.57, 0.36, 0.51}
\definecolor{DarkGreen}{rgb}{0.57, 0.8, 1}

\DeclareMathOperator{\diag}{diag}

\DeclareMathOperator{\rank}{rank}
\DeclareMathOperator{\Span}{span}
\DeclareMathOperator{\Null}{null}

\DeclareMathOperator{\blkdiag}{blkdiag}
\newtheorem{thm}{Theorem}
\newtheorem{prop}{Proposition}

\newtheorem{lem}{Lemma}
\newtheorem{assumption}{Assumption}
\newtheorem{remark}{Remark}
\newtheorem{definition}{Definition}

\title{\LARGE \bf
	{Observer-Based Control of Second-Order Multi-vehicle Systems in Bearing-Persistently Exciting Formations
	}}

\author{Zhiqi~Tang$^{1}$, Baris Fidan$^{2}$, Karl H. Johansson$^{1}$, Jonas Mårtensson$^{1}$,  and Tarek Hamel$^{3}$ 
	\thanks{}
	\thanks{$^1$ Division of Decision and Control Systems, School of Electrical Engineering and Computer Science, KTH Royal Institute of Technology, SE-10044 Stockholm, Sweden. {\tt\small ztang2@kth.se,kallej@kth.se,jonas1@kth.se}}
	\thanks{$^{2}$ Mechanical and Mechatronics Engineering Department, University of Waterloo, ON, Canada. {\tt\small fidan@uwaterloo.ca}}
	\thanks{$^{3}$ I3S-UCA, CNRS, Universit{\'e} C{\^o}te d'Azur $\&$ Institut Universitaire de France (IUF), France. {\tt\small thamel@i3s.unice.fr}}
}

\begin{document}
	
	\maketitle
	\cfoot{\thepage}
	\begin{abstract}
		
		This paper proposes an observer-based formation tracking control approach for multi-vehicle systems with second-order motion dynamics, assuming that vehicles' relative or global position and velocity measurements are unavailable.  It is assumed that all vehicles are equipped with sensors capable of sensing the bearings relative to neighboring vehicles and only one leader vehicle has access to its global position. Each vehicle estimates its absolute position and velocity using relative bearing measurements and the estimates of neighboring vehicles received over a communication network. A distributed observer-based controller is designed, relying only on bearing and acceleration measurements.
		This work further explores the concept of the \textit{Bearing Persistently Exciting} (BPE) formation by proposing new algorithms for bearing-based localization and state estimation of second-order systems in centralized and decentralized manners. It also examines conditions on the desired formation to guarantee the exponential stability of distributed observer-based formation tracking controllers. In support of our theoretical results, some simulation results are presented to illustrate the performance of the proposed observers as well as the observer-based tracking controllers. 
		
	\end{abstract}
	
	\vspace{0.3cm}
	
	\section{INTRODUCTION}
	Multi-vehicle systems are in demand to accomplish missions in different challenging scenarios, such as infrastructure inspection, surveillance, precision agriculture, exploration of deep waters, land, and space, etc. \cite{dorigo2020reflections}.
	During these coordination tasks, a multi-vehicle system must always be able to localize the position, estimate the velocity, and track desired trajectories in a decentralized fashion.
	
	Early works include localization based on Global Navigation Satellite Systems (GNSS). Due to the unreliability of GNSS in indoor and congested environments, dedicated onboard local sensors such as cameras are preferred to fulfill high precision requirements, which are passive,  lightweight, and power efficient. Vehicles equipped with onboard cameras can measure their neighboring vehicles' relative bearing (direction) measures if they are in the cameras' field of view \cite{franchi2012modeling}. Since bearing measurements are robust to noise and can be easily measured by cameras, bearing-based localization and control problems have become a popular research topic in recent years.
	
	Early work on bearing-based localization \cite{zhao2016localizability, van2020pose} relies on the bearing rigidity theory (also termed parallel rigidity  \cite{eren2003sensor}), which explores the conditions on graph topology of a multi-agent system that ensures the unique formation's configuration up to a scaling and a translational factor using constant inter-agent bearing measurements. More recent works present generalized bearing-based localization solutions that involve time-varying bearing measurements. The authors in \cite{tang2022relaxed} propose the concept of \textit{Bearing Persistently Exciting} (BPE) formation, which explores the uniqueness of the formation's configuration up only to a translational factor under a much-relaxed graph topology than bearing rigid formation, provided that a set of inter-agent bearings with the formation are persistently exciting (PE). However, existing works in the literature focus only on localization \cite{zhao2016localizability,tang2022localization}, that is, only on estimating the positions of each agent in the multi-vehicle systems. For multi-agent coordination involving fast vehicle dynamics (e.g., vehicles with second-order systems), the velocity of each vehicle also needs to be estimated. To the best of our knowledge, the bearing-based localization and velocity estimation problem for multi-agent systems with second-order agent motion dynamics still remains an open problem.
	

	Motivated by the above open problems, this paper proposes an observer-based formation tracking control approach for second-order $n (\geq 2)$-vehicle systems defined in $d (\geq 2)$-dimensional space. It is assumed that all vehicles are equipped with sensors capable of sensing their bearing relative to neighboring vehicles and that at least one leader vehicle has access to its global position. With the help of proposed optimal state
	observers, each vehicle estimates its absolute position and velocity using relative bearing measurements and the position estimates of neighboring vehicles. Both centralized and decentralized position and velocity observers are designed for the multi-vehicle system. The centralized observers are first proposed using a centralized Riccati gain. For the more efficient decentralized observers, a cascaded structure is adopted in the design: i) the first level is the design of individual Riccati observers to estimate relative position and velocity between neighboring vehicles whose inter-agent bearings are PE; ii) then the distributed observers for the $n$-agent systems is achieved using a Luenberger-type observer. We prove that estimation error is globally exponentially stable under both centralized and decentralized observers, provided the current formation is BPE. Finally, we propose an observer-based controller relying solely on vehicles' estimated positions and velocities and prove the local exponential stability of the decentralized observer-based tracking controller, provided that the desired formation is BPE.

	The body of the paper consists of seven sections. Section II provides preliminary knowledge about graph theory, the persistency of excitation, and bearing rigidity and bearing persistently exciting formation. Section III formulates the problem of this paper. Section IV proposes the designs for bearing-based localization and velocity estimation with stability analysis. Section V presents the integrated observer-based formation tracking controller together with stability analysis. Simulation results are presented in Section VI. The paper concludes with some final comments in Section VII.
	
	\section{Preliminaries}
	We denote by $S^{d-1}:=\{y\in\mathbb{R}^d:\|y\|=1\}$ the $(d-1)$-Sphere $(d\ge 2)$ and $\|.\|$ the Euclidean norm. 
	The matrix's null space and rank are denoted by $\Null(.)$ and $\rank(.)$, respectively. The set of symmetric positive-definite matrices of dimension $d\times d$  is denoted $\mathbb{S}_{+}(d)$. The matrix $I_d$ represents the identity matrix of dimension $d\times d$ and $\boldsymbol 1_n = [1,\ldots,1]^T \in \mathbb R^{n}$ the column vector of ones. The operator $\otimes$ denotes the Kronecker product and $\diag(A_i) = \blkdiag \{A_1, \ldots, A_n\} \in \mathbb{R}^{nd\times nd}$ the block diagonal matrix with elements given by $A_i\in \mathbb{R}^{d\times d}$ for $i=1,\ldots,n$.
	For any $y\in S^{d-1}$, we define the projection operator $\pi_y$
	\begin{align*}
		\pi_y := I - y y^{\top} \geq 0, 
	\end{align*}
	which projects any vector $x\in\mathbb{R}^d$ to the plane orthogonal to $y$.
	%
	%
	\subsection{Persistence of excitation}
	The following definition and persistence of excitation properties are borrowed from \cite{le2017observers}.
	
	\begin{definition}\label{def:peMatrix}
		A positive semi-definite matrix function $\Sigma:\mathbb R_{\ge 0}\to \mathbb{R}^{n\times n}$, satisfies \textit{persistently exciting} (PE) condition if there exists $T>0$ and $\mu>0$ such that for all $t\ge0$
		\begin{equation}
			\frac 1 T\int_{t}^{t+T}\Sigma(\tau)d\tau\ge\mu I. \label{eq:pe}
		\end{equation}
		\label{def:pe of matrix}
	\end{definition}
	\begin{definition}\label{def:pe}
		A direction $y(t)\in {S}^2$, is called \textit{persistently exciting} (PE) if the matrix function $\pi_{y(t)}$ satisfied the PE condition in Definition \ref{def:pe of matrix}.
	\end{definition}
	\begin{lem}\label{lem:Q_norm}
		Let $Q:=\sum_{i=1}^{l}\pi_{y_i}$. Then the matrix $Q$  satisfies the PE condition in Definition \ref{def:pe of matrix}, if one of the following conditions is satisfied:
		\begin{enumerate}
			\item there is at least one of the directions $y_i$ that is PE,
			\item there are at least two direction $y_{i}(t)$ and $y_j(t)$, $i,j\in\{1,2,...,l\},\ i\ne j$ that are persistently non-colinear, that is there exists $T>0$ and $\mu>0$ such that for all $t\ge0$,  $\frac 1 T \int_t^{t+T}(1-|y_i(\tau)^\top y_j(\tau)|)d\tau \geq \mu$.
		\end{enumerate}
	\end{lem}
	\subsection{Graph Theory}\label{subsec:graph}
	The interaction topology of a $n$-agent system can be modeled as an undirected graph $\mathcal{G} := (\mathcal{V}, \mathcal{E})$, where $\mathcal{V}=\{1,\ldots,n\}$ ($n \geq 2$) is the set of vertices and $\mathcal{E} \subseteq \mathcal{V} \times \mathcal{V}$ is the set of undirected edges. Two vertices $i$ and $j$ are called adjacent (or neighbors)
	when $(i,j)\in \mathcal{E}$. The set of neighbors of agent $i$ is denoted by $\mathcal{N}_i:=\{j\in\mathcal{V}|(i,j)\in\mathcal{E}\}$. If $j\in\mathcal N_i$, it follows that $i\in \mathcal N_j$, since the edge set in an undirected graph consists of unordered vertex pairs. $m=|\mathcal E|$ denotes the cardinality of the set $\mathcal E$. 
	The graph Laplacian matrix is defined as
	\begin{equation} \label{eq:Laplacian}\scalebox{0.95}{$
			L:=\bar H^\top  \bar H, \text{ with } \bar H=H \otimes I_d$}
	\end{equation}
	where $H$ is the incidence matrix and $\Null(L)=\Span(U)$ with $U=\boldsymbol 1 \otimes I_d$. If the graph is connected, one has $\rank(L)=\rank(\bar H)=dn-d$.
	
	%


\subsection{Bearing Rigidity and Bearing Persistently Exciting Formation}
Consider a group of $n$ agents, where the position of each agent $i, i\in\{1,\ldots,n\}$,  expressed in an inertial frame common is denoted as $p_i\in \mathbb R^{d}$. Assume the interaction topology between the group of agents is an undirected graph $\mathcal G$, then
the graph $\mathcal{G}$ together with the configuration $\boldsymbol{p}:=[p_1^\top,...,p_n^\top]^\top\in \mathbb{R}^{dn}$ define formation  $\mathcal{G}(\boldsymbol p)$ in the $d$-dimensional space.

Define the relative position vectors
\begin{equation}
	\label{eq:eij}
	p_{ij}:=p_{j}-p_i, \ (i, j) \in \mathcal{E}
\end{equation}
and if $\|p_{ij}\|\neq 0$, the bearing measurement of agent $j$ relative to agent $i$ is given by the unit vector
\begin{equation}
	\label{eq:gij}
	g_{ij} := p_{ij}/\|p_{ij}\| \in S^{d-1}.
\end{equation}

We call a group of agents 'bearing formation', if each agent $i$ can sense the relative bearings $g_{ij}$ to its neighboring agents $j\in \mathcal N_i$.
To analyze the localizability of a bearing formation,  existing works in the literature have introduced the concept of the bearing Laplacian matrix which is defined as 
\begin{equation}\label{eq:LB}\scalebox{0.95}{$
		L_B:=\bar H^\top \Pi \bar H \text{, with } \Pi=\diag(\pi_{\bar g_k}).$}
\end{equation}
where $$\bar g_k:= \frac{\bar p_k}{\|\bar p_k\|}\in S^{d-1},\ k\in \{1,2,\ldots,m\}$$
and 
\begin{equation}\label{eq:p_k}
	\bar p_k:= p_{ij},\ k\in \{1,2,\ldots,m\},
\end{equation}
denote the edge vector with assigned direction under an arbitrary orientation of the graph, such that $i$ and $j$ are, respectively, the initial and the terminal nodes of $\bar p_k$. 
Since $\Span\{U,\boldsymbol p\}\subseteq \Null(L_B(\boldsymbol p))$,  it follows that $\rank(L_B)\le dn-d-1$. 

Existing works mainly focus on estimating the formation's position configuration using relative bearing measurements between neighboring agents and each agent's velocity input measures. In these works (e.g., \cite{zhao2016localizability,tang2022relaxed}), each agent is typically modeled as a single-integrator system. The definitions used for bearing-based localization problems are recalled below.

A formation is called Infinitesimal Bearing Rigid (IBR) if its position configuration can be uniquely determined using constant bearing measurements up to a translation and a scaling factor. In this situation, one has $\rank(L_B)= dn-d-1$ and $\Null(L_B(\boldsymbol p))=\Span\{U,\boldsymbol p\}$ for each fixed configuration $\boldsymbol p$. 

In contrast to IBR formation, BPE formation is a time-varying bearing formation introduced first in \cite{tang2022relaxed}. If a formation is BPE, its position configuration can be uniquely determined up to a translational factor using relative bearing measurements and each agent's velocity input.  The definition of BPE formation under a fixed graph topology is recalled as follows: 

\begin{definition}\label{DefBPE}
	A formation $\mathcal G(p)$ is called BPE if $\mathcal G$ is connected and the bearing Laplacian matrix is persistently exciting (PE), i.e.,
	there exists $ T>0, \mu>0$ such that  $\forall t\ge0$
	\begin{equation}
		\scalebox{0.9}{$
			\begin{aligned}
				\frac 1 T \int_{t}^{t+T}L_B(\boldsymbol p(\tau))d\tau=\frac 1 T {\bar H}^\top\int_{t}^{t+T} {\Pi}(\boldsymbol p(\tau))d\tau{\bar H}\ge\mu L. \label{eq:pe_LB}
			\end{aligned}$}
	\end{equation}
\end{definition}

		
		Note that the above PE condition for the bearing Laplacian is less restrictive than the PE condition from Definition 1. In particular, having a matrix, $ \Pi$, that is PE is sufficient but not necessary to ensure \eqref{eq:pe_LB}. 
		
		It is useful to remember that the concept of BPE formation implies that some inter-agent bearings are time-varying (and PE).
		The next Lemma provides a condition on the number of PE bearings that is necessary for having a BPE formation 
		when $(n-1)<m < \bar{m}$, that is, the number of edges $m$ is below the minimal number $\bar m$ of edges required to have a IBR formation. 
		
		\begin{figure}[!t]
			\centering
			\includegraphics[width=2in]{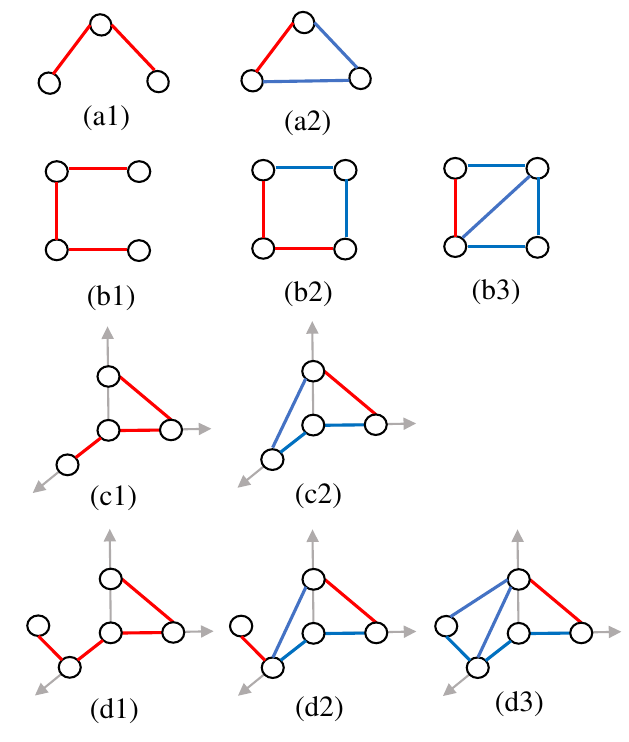}	
			\caption{Eamples of BPE formations in two (a1-b3) and three-dimensional space (c1-d3). Red lines represent edges for which the corresponding bearing vectors are PE, and blue lines represent edges for which the corresponding bearing vectors are not necessarily PE. }
			\label{fig:pe_edges3D}
		\end{figure}
		
		\begin{lem}\label{lem:necessary}
			Consider a BPE formation $\mathcal G (\boldsymbol p(t))$ defined in $\mathbb R^{d}$ with $n$ agents and $m$ edges, where $(n-1)\le m <\bar{m}$,
			then the number of PE bearing vectors inside the formation, $m_{PE}$, satisfies  $ m_{PE} \geq d(n-1)-(d-1)m$.
		\end{lem}
		The proof of this Lemma can be found in \cite{tang2022some} and the graphical examples are shown in Fig. \ref{fig:pe_edges3D}.

		\section{Problem formulation}
		Consider a group of $n$ vehicles with the double integrator motion dynamics model
		\begin{equation}\label{eq:double integrator}
			\left\{
			\begin{aligned}
				\dot{p}_i&=v_i\\
				\dot{v}_i&=u_i,\ i=1,...,n.
			\end{aligned}
			\right.
		\end{equation}
		where $p_i\in \mathbb R^d$ and $v_i\in \mathbb R^d$ are the position and velocity, respectively, of the $i$th vehicle expressed in a common inertial frame. $u_i\in \mathbb R^d$ denotes as the acceleration control input. These vehicles are required to move in a formation in a $d$-dimensional space under the following standing hypothesis.
		\begin{assumption}\label{ass:leader}
			At least one agent $i\in\{1,\dots, n\}$ in the formation can measure its own position $p_i$ and velocity $v_i$. Such an agent is referred to as the leader.
		\end{assumption}
		
		The global positioning information is not available for follower agents. However, relative bearing measurements are obtained through onboard sensors. More precisely, if the $i$th vehicle that “sees” the neighbor agent $j\in \mathcal N_i$, it can measure the relative bearing $g_{ij}$.
		Then, it is assumed that a bidirectional interconnection exists between adjacent nodes, as described below.
		\begin{assumption} \label{ass:construction}
			The topology of the group is described by a graph topology $\mathcal G$ defined in Subsection \ref{subsec:graph}. Each agent $i\in\mathcal V$ can measure the relative bearing vectors $g_{ij}$ to its neighbors $ j\in \mathcal{N}_i$  in a common inertial frame. Besides, each agent can transmit its state estimates to its neighbor agents through communication.
		\end{assumption}
		
		Then, the bearing-based localization and formation tracking control problem is
		posed as follows. 
		By assigning one leader in the formation, 1) design localization algorithms using acceleration input of each agent and the inter-agent bearings to estimate the agents' positions and velocities and 2) design distributed feedback controllers using the estimated positions and velocities as feedback information to asymptotically track any feasible desired formation.

		\section{Bearing-based localization and velocity estimation}

			By assigning one leader in the formation knowing its own position, we propose centralized and decentralized localization algorithms to estimate the position and velocity of each follower using only the bearings and acceleration input of each agent.  
			Without loss of generality, we consider a formation in which agent $1$ is the leader, the unique agent in the formation that measures its own position $p_1$. The other (follower) agents can measure only relative bearings to their neighboring agents. 
			\begin{assumption}\label{ass:current}
				For all agents $i\in\mathcal V$ and $\forall t$, the velocities $v_i(t)$ and positions $p_i(t)$, and acceleration input $u_i$ are bounded, the resulting relative bearing measurements $g_{ij}(t)$ are well-defined and the formation $\mathcal G(\boldsymbol p(t))$  is BPE.
			\end{assumption}
			Using the above description along with \eqref{eq:double integrator}, one describes the systems dynamics as follows:
			\begin{equation}\label{eq:n-systems}
				\begin{bmatrix} \dot {\boldsymbol p} \\ \dot {\boldsymbol v}\end{bmatrix}=A \begin{bmatrix} \boldsymbol p \\ \boldsymbol v\end{bmatrix}+B\boldsymbol u, \quad \boldsymbol y=C \begin{bmatrix} \boldsymbol p \\  \boldsymbol v\end{bmatrix}
			\end{equation} 
			where $\boldsymbol v=[v_1^\top,\ldots, v_n^\top]^\top$ and $\boldsymbol u=[u_1^\top,\ldots, u_n^\top]^\top$, $A=\begin{bmatrix}
				0_{dn} & I_{dn}\\ 0_{dn} & 0_{dn} 
			\end{bmatrix}$,  $B=\begin{bmatrix} 0_{dn} \\I_{dn} \end{bmatrix}$ and $C=\begin{bmatrix}L_B(\boldsymbol p)+C_1 & 0 \end{bmatrix}\in \mathbb R^{dn\times 2dn}$, with $C_1=\diag(I_d, 0_d, \ldots,0_d)$. 
			
			Note that when there is no leader $(C_1 \equiv 0_{nd})$, the output $\boldsymbol y$ becomes an implicit output; that is $\boldsymbol y=C \begin{bmatrix} \boldsymbol p  \\  \boldsymbol v\end{bmatrix}=0$ due to the fact of $L_B(\boldsymbol p) \boldsymbol p\equiv 0$.  Let $\hat p_i \in \mathbb{R}^{d}$ and $\hat v_i \in \mathbb{R}^{d}$  denote the estimate of $p_i$ and $v_i$, respectively. 
			
			\begin{thm}[Centralized state estimation]\label{thm:centralize}
				Consider the $n$-agent system $\mathcal G (\boldsymbol p(t))$ defined by \eqref{eq:n-systems}. Define stacked position and velocity estimates as $\hat {\boldsymbol p}=[\hat p_1^\top,\hat p_2^\top,\ldots,\hat p_n^\top]^\top$ and $\hat {\boldsymbol v}=[\hat v_1^\top,\hat v_2^\top,\ldots,\hat v_n^\top]^\top$ along with the following dynamics: 
				\begin{equation} \label{eq:observer}
					\begin{bmatrix} \dot {\hat{\boldsymbol p}} \\ \dot {\hat{\boldsymbol v}}\end{bmatrix}=A \begin{bmatrix} \hat{\boldsymbol p} \\ \hat{\boldsymbol v}\end{bmatrix}+B\boldsymbol u+K\left (\boldsymbol y-C\begin{bmatrix} \hat{\boldsymbol p} \\ \hat{\boldsymbol v}\end{bmatrix}\right)
				\end{equation}
				with $K(t)=\kappa M(t)C^\top Q \in \mathbb R^{2dn\times dn}$, $\kappa \geq \frac{1}{2}$ and $M(t)$ solution of the following Continous Riccati Equation (CRE):
				\begin{equation}\label{riccati}
					\dot M=AM+MA^\top -MC^\top Q C M+S,\quad M(0)=I_{2nd}
				\end{equation}
				$Q\in \mathbb{S}_{+}(dn)$ and $S\in \mathbb{S}_{+}(2dn)$ are positive definite matrices.
				Then, if Assumptions \ref{ass:leader}-\ref{ass:current} are fulfilled, the origin of the estimation error $\boldsymbol \delta=(\boldsymbol \delta_p^\top,\boldsymbol \delta_v^\top)^\top:=((\hat {\boldsymbol p}-\boldsymbol p)^\top, (\hat {\boldsymbol v}-\boldsymbol v)^\top)^\top$ is globally exponentially stable.
			\end{thm}

			\begin{proof}
				Consider the following candidate Lyapunov function:  
				\[{\cal L}(\boldsymbol \delta,t):=\boldsymbol \delta^\top  M^{-1}(t)\boldsymbol \delta. \] 
				From \eqref{eq:n-systems}-\eqref{eq:observer} along with \eqref{riccati}, one deduces that:
				\begin{align*}
					\dot{{\cal L}}=&-\boldsymbol \delta^\top\big((2\kappa-1)C^\top  QC+M^{-1}SM^{-1}\big)\boldsymbol \delta, \end{align*}
				is negative definite as long as $M$ is positive definite and well-conditioned. That is, there exist two positive numbers $p_{m}$, and $p_M$, such that $p_m I_{nd} \leq M(t) \leq p_M I_{nd}$.
				Since $A$ is a real nilpotent matrix, one concludes that all its eigenvalues are zero. Using the fact that $C=\begin{bmatrix}L_B(\boldsymbol p)+C_1 & 0_{dn} \end{bmatrix}\in \mathbb R^{dn\times 2dn}=(L_B(\boldsymbol p)+C_1)\mathring C$, with $\mathring C=\begin{bmatrix} I_{dn} & 0_{dn} \end{bmatrix}$ one deduces that the pair $(A,\mathring C)$ is Kalman observable. This implies, using Lemma \ref{extension} (see the Appendix), that the equilibrium $\boldsymbol \delta=0$ is uniformly observable if $ L_B(\boldsymbol p)+C_1$ is satisfies the PE condition in Definition \ref{def:pe of matrix}. From there, one deduces that the Riccati solution \eqref{riccati} is well conditioned, and hence, one concludes that  ${\cal L}$ converges exponentially to zero.
				
			\end{proof}
			\subsection{Decentralized localization and velocity estimation}
			We propose a cascade structure to design an efficient decentralized observer. The first level focuses on designing decentralized observers for relative positions and velocities associated with the $ m_{PE}$ persistently exciting bearings by deriving $m_{PE}$ reduced Riccati observers of dimension $2d+d(2d+1)$ states each\footnote{$2d$ for the relative state $(\bar{p}_k,\bar{v}_k)$ and $d(2d+1)$ for the associated symmetric Riccati matrix $M_k$.}. The second level is a decentralized Luenberger-type observer of dimension $2dn$, achieved afterward.  This structure brings simplicity and robustness; the computational load is more manageable (linear on $n$); and finally, any bearing formation with a large number of agents can be quickly processed compared to the centralized structure that requires $2dn +dn(2dn+1)$ states in total. 
			
			Let $\mathcal S_e$ be the set containing all PE inter-agent bearing measurements. For each relative variables $(\bar p_k,\bar v_k)$ associated to PE bearing $\bar g_k\in \mathcal S_e$, one has:
			\begin{equation}\label{relative}
				\begin{bmatrix} \dot {\bar p}_k \\ \dot {\bar v}_k\end{bmatrix}=A_k \begin{bmatrix} {\bar p}_k \\ {\bar v}_k\end{bmatrix}+B_k \bar u_k, \quad \bar y_k=C_k \bar p_k
			\end{equation}
			with $\bar v_k:=\dot {\bar p}_k=v_i-v_j$, $\bar u_k:=\dot {\bar v}_k=u_i-u_j$, $A_k=\begin{bmatrix}
				0_{d} & I_{d}\\ 0_{d} & 0_{d} 
			\end{bmatrix}$, $B_k=\begin{bmatrix} 0_{dn} \\I_{dn} \end{bmatrix}$ and $C_k=\pi_{\bar g_k}\begin{bmatrix}I_d & 0_d\end{bmatrix}$.
			
			The first step of the proposed observer focuses only on the relative dynamics. 
			\begin{lem}\label{lem:estimator_ij}
				Consider the relative dynamics \eqref{relative} associated with a PE bearing $\bar g_{k} \in \mathcal S_e $. Then, the following relative state observer: 
				\begin{equation}\label{eq:observer_ij}
					\begin{bmatrix} \dot {\hat{\bar p}}_k \\ \dot {\hat{\bar v}}_k\end{bmatrix}=A_k \begin{bmatrix} \hat{\bar p}_k \\ \hat{\bar v}_k\end{bmatrix}+B_k\bar u_k-K_k C_k\begin{bmatrix}\hat{\bar p}_k \\ \hat{\bar v}_k\end{bmatrix}
				\end{equation}
				with $K_{k}=\kappa_{k}M_kC_k^\top Q_k\in \mathbb R^{2d\times d}$, $\kappa_k\geq \frac{1}{2}$, $M_k(t)$ solution to the Continuous Riccati Equation (CRE)
				\begin{equation}\label{relativeM}
					\dot M_k=A_kM_k+M_kA_i^\top -M_kC_k^\top Q_k C_k M_k+S_k,
				\end{equation}
				and $Q_k\in \mathbb S_{+}(d)$ and $V_k \in \mathbb S_{+}(2d)$, 
				globally exponentially stabilizes the estimation error $ \bar \delta_k=(\hat{\bar p}_k-\bar p_k, \hat{\bar v}_k-\bar v_k)$  to 0. 
			\end{lem}
			\begin{proof}
				Analogously to the proof Theorem \ref{thm:centralize}, one considers the following candidate Lyapunov function,
				\begin{equation}\label{M_k}
					{\cal L}_{\bar \delta_k}:=\bar \delta_k^\top  M_k^{-1}(t)\bar \delta_k.
				\end{equation}
				with 
				\begin{align}\label{dot_M_k}
					\dot{{\cal L}}_{\bar \delta_k}=&-\bar \delta_k^\top\big((2\kappa_k-1)C_k^\top  Q_kC_k+M_k^{-1}S_kM_k^{-1}\big)\bar \delta_k, \end{align}
				The remaining part of the proof is similar to the proof of Theorem \ref{thm:centralize}. It is omitted here for conciseness. 
			\end{proof}
			
			For the second step, we define the pseudo-'bearing Laplacian matrix' $ \bar L_B=\bar H^\top \bar \Sigma \bar H$ by setting $\bar \Sigma=\diag(\sigma_1, \ldots, \sigma_m)$, with $\sigma_k=I_d$ if the bearing $\bar g_k$ is PE  and $\sigma_k=\pi_{\bar g_k}$ otherwise, $\forall k \in \{1,\ldots,m\}$.
			\begin{lem}\label{lem:LB_bar}
				If $\mathcal G(p)$ is BPE according to Definition \ref{DefBPE}, and the set of non-PE bearings is empty or composed of constant bearing, then the pseudo-bearing Laplacian matrix $\bar{L}_B$ verifies that 
				there exists $ T>0, \mu>0$ such that  $\forall t\ge0$
				\begin{equation}\scalebox{0.9}{$
						\begin{aligned}
							\bar{L}_B \geq \frac 1 T \int_{t}^{t+T}L_B(\boldsymbol p(\tau))d\tau=\frac 1 T {\bar H}^\top\int_{t}^{t+T} {\Pi}(\boldsymbol p(\tau))d\tau{\bar H}\ge\mu L. \label{eq:pe_LB_}
						\end{aligned}$}
				\end{equation}
				Moreover, $\bar L_B+C_1$ is an invertible constant matrix. 
			\end{lem}
			\begin{proof}
				Since $I_d \geq \sigma_k$, it is straightforward to verify that $\bar{L}_B \geq L_B ={\bar H}^\top {\Pi}(\boldsymbol p){\bar H}$. From there and using the fact that $\bar L_B$ is constant, one gets \eqref{eq:pe_LB_}. 
				Now, since $\bar L_B\geq \mu L$, $\rank(L)=dn-d$, $\rank(C_1)=d$, and $\ker(C_1)$ is uniformly non-colinear to $\ker(L)$ (resp. $\ker(\bar L_B$)), one ensures that $\bar L_B+C_1$ is invertible.  
			\end{proof}
			
			Define $\hat{\bar{\boldsymbol p}}=[\hat{\bar p}_1^\top,\hat{\bar{p}}_2^\top\ldots,\hat{\bar{p}}_k^\top,\ldots, \hat{\bar p}_m^\top]^\top$ (respectively $\hat{\bar{\boldsymbol v}}=[\hat{\bar v}_1^\top,\hat{\bar{v}}_2^\top\ldots,\hat{\bar{v}}_k^\top,\ldots, \hat{\bar v}_m^\top]^\top$) where $\hat{\bar p}_k$ (respectively $\hat{\bar v}_k$) is obtained from the integration of \eqref{eq:observer_ij}-\eqref{relativeM} if $\bar g_k$ is PE and $(\hat{\bar p}_k, \hat{\bar v}_k)$ is set to zero ($(\hat{\bar p}_k, \hat{\bar v}_k)=(0,0)$) otherwise. 
			
			\begin{thm}\label{TH-dec}Consider a $n$-agent system $\mathcal G (\boldsymbol p(t))$ defined in $\mathbb R ^{d}$. Under assumptions \ref{ass:leader}-\ref{ass:current} along with the statement of Lemma \ref{lem:estimator_ij} and \ref{lem:LB_bar}, and the following distributed observer:
				\begin{equation} \label{eq:dist_observer}
					\begin{bmatrix} \dot {\hat{\boldsymbol p}} \\ \dot {\hat{\boldsymbol v}}\end{bmatrix}=A \begin{bmatrix} \hat{\boldsymbol p} \\ \hat{\boldsymbol v}\end{bmatrix}+B\boldsymbol u+\Lambda\left(\bar H^\top \bar \Sigma(\hat {\bar{\boldsymbol p}}-\bar H \hat{\boldsymbol p} )+C_1(\boldsymbol p-\hat{\boldsymbol p})\right)
				\end{equation}
				with $\Lambda=\begin{bmatrix}\kappa_{o_1}I_{dn}\\\kappa_{o_2}I_{dn}\end{bmatrix}$ and $\kappa_{o_1}$ and $\kappa_{o_2}$ positive gains, the origin of the estimation error $(\boldsymbol \delta_{p},\boldsymbol \delta_{ v})$ is globally exponentially stable.
			\end{thm}
			\begin{proof}
				To show the distributed nature of the proposed estimator \eqref{eq:dist_observer}, it is straightforward to verify that, for each agent, one has: 
				\begin{equation}\label{eq:estimator_pi}
					\left\{\begin{aligned}\dot{\hat p}_i&=\hat v_i+\kappa_{o_1}\sum_{j\in \mathcal N_i}\sigma_{ij}\left( \hat{p}_{ij}-(\hat p_i-\hat p_j)\right)\\
						\dot{\hat v}_i&=u_i+\kappa_{o_2}\sum_{j\in \mathcal N_i}\sigma_{ij}\left( \hat p_{ij}-(\hat p_i-\hat p_j)\right), i\in\mathcal V/\{1\}
					\end{aligned}\right.
				\end{equation}
				and:
				\begin{equation}\label{eq:estimator_p1}
					\left\{\begin{aligned}\dot{\hat p}_1&=\hat v_1+\kappa_{o_1}\sum_{j\in \mathcal N_1}\sigma_{1j}\left( \hat p_{ij}-(\hat p_1-\hat p_j)\right)-(\hat p_1-p_1)\\
						\dot{\hat v}_1&=u_1+\kappa_{o_2}\sum_{j\in \mathcal N_1}\sigma_{1j}\left( \hat p_{ij}-(\hat p_1-\hat p_j)\right)-(\hat p_1-p_1)
					\end{aligned}\right.
				\end{equation}
				Note that $\hat p_{ij}\neq \hat p_j-\hat p_i$. It denotes the estimate of edge vector $\hat p_{ij}:=\hat {\bar p}_k$  defined by \eqref{eq:p_k} using \eqref{relative}-\eqref{relativeM} if the associated bearing is PE ($\hat p_{ij}=0$ if the bearing $g_{ij}$ is not PE). 
				
				Using \eqref{eq:n-systems} and \eqref{eq:dist_observer}
				along with the fact that $\pi_{g_{ij}}p_{ij}= 0$ and hence $\bar H^\top \bar \Sigma \bar{\boldsymbol p}=\bar H^\top \bar \Sigma \bar H \boldsymbol p$ with $\bar{\boldsymbol p}=[{\bar p}_1^\top,{\bar p}_2^\top\ldots,{\bar p}_k^\top,\ldots, {\bar p}_m^\top]^\top$, one has
				\begin{align} \label{eq:delta}
					\dot{\boldsymbol \delta}&=A \boldsymbol \delta-\Lambda\left((\bar L_B+C_1)\boldsymbol \delta_p-\bar H^\top \bar \Sigma \mathring{C}\bar{\boldsymbol \delta}\right) \notag\\
					&=(A-\Lambda\bar C) \boldsymbol \delta+\Lambda \bar H^\top \bar \Sigma \mathring{C}\bar{\boldsymbol \delta}
				\end{align}
				with $\bar{\boldsymbol \delta}=( \bar{\boldsymbol \delta}_p^\top, \bar{\boldsymbol \delta}_v^\top)^\top:=((
				\hat{\bar{\boldsymbol p}}-\bar{\boldsymbol p})^\top, (
				\hat{\bar{\boldsymbol v}}-\bar{\boldsymbol v})^\top)^\top$, $\bar C=(\bar L_B+C_1)\mathring{C}$, and $\mathring C=[I_{dn} \ 0_{dn}]$. 
				
				Since the error $\bar{\boldsymbol \delta}=0$ is exponentially  stable (from Lemma \ref{lem:estimator_ij}), one gets the following nominal dynamics: 
				\begin{equation} \label{eq:delta_unforce}
					\dot{\boldsymbol \delta}=(A-\Lambda\bar C) \boldsymbol \delta
				\end{equation}
				Using the fact that the system $(A,\mathring{C})$ is Kalman observable and $(\bar L_B+C_1)$ is invertible, one concludes that $(A,\bar{C})$ is also Kalman observable. Since \eqref{eq:delta_unforce} represents the dynamics of a second-order block-based system, then for any gain $\Lambda$ with positive entries $(\kappa_{o_1},\kappa_{o_2})$, there exist two constant positive definite matrices $P_{\boldsymbol \delta}$ and $Q_{\boldsymbol \delta} \in \mathbb S_+ (2nd)$ such that:
				\[\frac{d}{dt} \boldsymbol \delta^\top P_{\boldsymbol \delta} \boldsymbol \delta=-\boldsymbol \delta^\top Q_{\boldsymbol \delta} \boldsymbol \delta.\]

				From there, one concludes that $\boldsymbol \delta=0$ is also globally exponentially stable. 
				
			\end{proof}
			\begin{remark}
				In the particular case when all the bearings are
				constant, Theorem \ref{TH-dec} still holds, provided that the measures of relative positions are used instead of the estimated ones. Theorem \ref{TH-dec} directly encompasses IBR formations involving two adjacent leaders knowing their positions. It also applies to an IBR formation with non-adjacent leaders by adequately adjusting the expression of $C_1$. 
			\end{remark}
			\section{Observer-based control}
			This section is devoted to a distributed control design for the formation tracking problem. Although feedback information can be obtained from the centralized or decentralized cascaded observer, our primary focus will be on the decentralized cascaded observer-based controller.
			
			\begin{assumption}\label{ass:desired}
				The desired accelerations of each agent $u_i(t)^*$ is uniformly continuous and bounded such that the desired $v_i^*(t):=\dot p^*(t)$ and desired position $p_i^*(t)$ ($i\in\mathcal V$) are bounded, the resulting desired bearings $g_{ij}^*(t)$ are well-defined and the desired formation $\mathcal G(\boldsymbol p^*(t))$  is BPE.
			\end{assumption}
			\begin{prop}
				Consider a $n$-agent system $\mathcal G (\boldsymbol p(t))$ defined in $\mathbb R ^{d}$. Assume that  Assumptions \ref{ass:leader}, \ref{ass:construction} and \ref{ass:desired} are satisfied. Then under the decentralized cascaded observer \eqref{eq:observer_ij}, \eqref{eq:estimator_pi}, \eqref{eq:estimator_p1}, the following controller
				\begin{equation}\label{eq:control}
					u_i=-\kappa_{p_i}(\hat p_i-p_i^*)-\kappa_{v_i}(\hat v_i-v_i^*)+u^*_i
				\end{equation}
				ensures that the actual formation $\mathcal G(\boldsymbol p(t))$  is BPE and the tracking error $(\boldsymbol p-\boldsymbol p^*, \boldsymbol v-\boldsymbol v^*)$  is locally exponentially stable. 
			\end{prop}
			\begin{proof}
				Define $\tilde{\boldsymbol x}=[\tilde{\boldsymbol p}^\top \ \tilde{\boldsymbol v}^\top]^\top$, with $\tilde{\boldsymbol p} =\boldsymbol p-\boldsymbol p^*$, and $\tilde{\boldsymbol v}= \boldsymbol v-\boldsymbol v^*$. Using \eqref{eq:n-systems} along with \eqref{eq:control}, one gets: 
				$$\dot{\tilde{\boldsymbol x}}=(A-B_x) \tilde{\boldsymbol x}-B_x\boldsymbol \delta.$$
				with $B_x=\begin{bmatrix} 0_{nd} & 0_{nd} \\ \diag(\kappa_{p_i}I_{d}) & \diag(\kappa_{v_i}I_{d})\end{bmatrix}$. 
				
				Once again, this is as a second-order block-based system involving positive gains $(\kappa_{p_i}, \kappa_{v_i})$. Hence, there exist two positive definite matrices $P_x$ and $Q_x\in \mathbb S_+ (2nd)$ , such that, if $\boldsymbol \delta\equiv 0$, one has:
				\[\frac{d}{dt} \tilde{\boldsymbol x}^\top P_{x} \tilde{\boldsymbol x}=-\tilde{\boldsymbol x}^\top Q_{x} \tilde{\boldsymbol x}.\]
				
			
			Define a candidate Lyapunov function for the closed-loop system:
			
			\begin{equation} \label{W}
				\mathcal W= \alpha\sum_{k=1}^m\delta_k^\top M_k^{-1}\delta_k+ \beta\boldsymbol \delta ^\top P_{\boldsymbol \delta} \boldsymbol \delta+ \tilde{\boldsymbol x}^\top P_x \tilde{\boldsymbol x}. 
			\end{equation}
			
			where $M_k$ is the solution of the Riccati equation \eqref{relativeM}, $P_{\boldsymbol \delta}$ and $P_x$ are constant  positive definite matrices introduces previously and $\alpha$ and $\beta$ are two positive constants specified later. 
			Differentiating \eqref{W}, one gets:
			
			\begin{align}
				\dot{\mathcal W}=&-\alpha\sum_{k=1}^m \delta_k^\top\big((2\kappa_k-1)C_k^\top  Q_kC_k+M_k^{-1}S_kM_k^{-1}\big)\delta_k \notag\\
				&-\beta\boldsymbol \delta^\top Q_{\boldsymbol \delta} \boldsymbol \delta-2\beta\boldsymbol \delta^\top\bar P\Lambda\bar H^\top \bar \Sigma \mathring{C}(\bar{\boldsymbol \delta}) \notag\\
				&-\tilde{\boldsymbol x}^\top Q_x\tilde{\boldsymbol x}-2\tilde{\boldsymbol x}^\top P_x B_x\boldsymbol \delta
			\end{align}
			Now, if  $\mathcal G(\boldsymbol p(t))$  is BPE, one ensures that there exist two positive numbers $p_{m}$, and $p_M$, such that $p_m I_{d} \leq M_k(t) \leq p_M I_{d}$, $\forall \bar g_k\in \mathcal S_e$ and therefore:
			\begin{align*}
				\dot{\mathcal W}\leq &-\alpha \frac{s_m p_m}{p_M^2}\sum_{k=1}^m \|\delta_k\|^2   -\beta q^\delta_m\|\boldsymbol \delta\|^2    -q^x_m\|\tilde{\boldsymbol x}\|^2\notag\\
				& \quad \quad \quad \quad \quad \quad \quad \quad \quad +2\beta \gamma \|\boldsymbol \delta\|\|\bar{\boldsymbol \delta}\| + 2\lambda \|\tilde{\boldsymbol x}\|\|\boldsymbol \delta\|
			\end{align*}
			with $s_M$, $q_m^\delta$, $q_m^x$ positive numbers such that:
			\[S_k \geq s_M I_{2d},\; Q_{\boldsymbol \delta} \geq q_m^\delta I_{2nd},\; Q_{x} \geq q_m^x I_{2nd} \]
			and $\gamma$ and $\lambda$ verifies:
			\[\gamma \geq \|\delta^\top\bar P\Lambda\bar H^\top \bar \Sigma \mathring{C}\|_2, \; \lambda \geq \|P_x B_x\|_2 \]
			
			From there, by choosing $\alpha$ and $\beta$ sufficiently large ($\alpha>\frac{\beta \gamma^2 p_M^2}{2 q_m^\delta p_m s_m}$ and $\beta >\frac{\lambda^2}{2 q_m^x q_m^\delta}$), one verifies that $\dot{\mathcal W}$ is negative definite, and hence one concludes that $\tilde{\boldsymbol x}$ is exponentially stable. 
			
			To prove that  $\mathcal G(\boldsymbol p(t))$  is BPE, we use a proof by contradiction. Assume there exists $k=\{1,\ldots, m\}$ for which $\bar{g}_k$ is not PE.
			
			Then, according to \cite{morin2017}, $\forall T>0$,:
			\[\exists \{t_p\}_{p\in \mathbb{N}}, \; \exists x \in S^{d-1}| \; \lim_{p \to \infty} \int_{t_p}^{t_p+T} x^\top\pi_{\bar g_k}(\tau)xd\tau=0.\]
			Since this should be valid for any $T>0$, by choosing $T$ as large as desired, one can consider ${\bar g_k}$ approximately constant. This, in turn, implies that \eqref{M_k} is no longer positive definite function since $\lim_{p \to \infty} \lambda_{\max}(M_k(t_p+T))\to \infty $. Assume that $\bar p_k$ does not cross zero, in order to ensure that $\bar g_k$ is well defined and define $\mathring{\bar{\delta}}_k=\diag(\pi_{\bar g_k},\pi_{\bar g_k})\delta_k$ and $\mathring{M}_k=\diag(\pi_{\bar g_k},\pi_{\bar g_k})M_k^{-1}\diag(\pi_{\bar g_k},\pi_{\bar g_k})$. Using $\mathring{\bar{\delta}}_k$ and  $\mathring{M}_k$ in  \eqref{M_k} one verifies  \eqref{dot_M_k} is still valid in this new variables ensuring that the reduced matrix $\mathring{M}_k/\{g_k\}$ is well-conditioned. From there, one ensures that $\diag(\pi_{\bar g_k})\delta_k \to 0$ and hence  $\pi_{\bar g_k}\hat{\bar p}_k=\pi_{\bar g_k}\hat{\bar v}_k=0$.
			Recalling now the fact that $p_{ij}:=\bar{p}_k$, $\hat{p}_{ij}:=\hat{\bar{p}}_k$, and $g_{ij}:=\bar g_k$, and using \eqref{relative} along with \eqref{eq:control}, one verifies that:
			\[\ddot p_{ij}=-\kappa_{p_i}(\hat p_{ij}-p_{ij}^*)-\kappa_{v_i}(\hat v_{ij}-v_{ij}^*)+u^*_{ij}\]
			Multiplying both sides of the above equation by $\pi_{g_{ij}}$, and exploiting the fact that $g_{ij}$ is in the same direction as $p_{ij}$ and $v_{ij}$ and consequently as $\ddot p_{ij}:=u_{ij}$, one gets:
			\[\pi_{g_{ij}} \left(\ddot{p}_{ij}^*+\kappa_{v_i}\dot{p}_{ij}^*+\kappa_{p_i}p_{ij}^*\right)=0,\]
			From this relationship one concludes that the desired trajectory of $p_{ij}^*$ (or equivalently $g_{ij}^*$) is in the direction of $g_{ij}$ and hence not PE which contradicts the assumption.
		\end{proof}
		\begin{remark}
			Note that the local exponential stability is the only conclusion drawn from the distributed observer-based tracking controller, as nothing prevents the relative position $\bar{p}_k$ from crossing zero at certain instances, thereby resulting in an ill-defined $\bar{g}_k$. 
		\end{remark}

		\section{Simulation Results}\label{sec:sim}
		\begin{figure}[!htb]
			\centering
			\includegraphics[scale = 0.5]{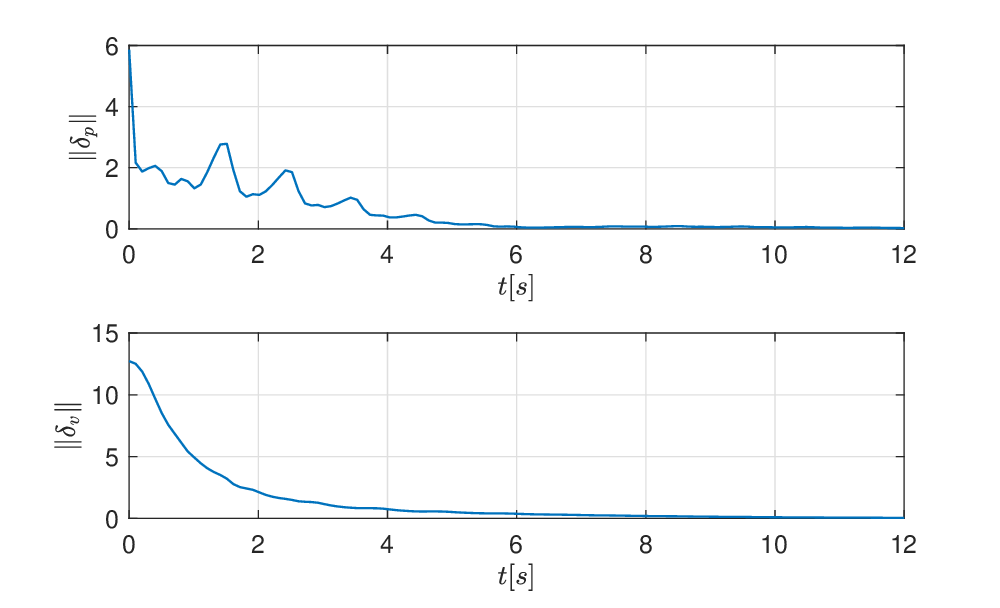}	
			\caption{Evolution of the estimation error $(\boldsymbol \delta_p,\boldsymbol \delta_v)$ under the centralized observer with measurement noises. Gains are chosen as: $\kappa=10,\ M(0)=I_{2dn},\ Q=10I_{dn}, \ S=0.01I_{2dn}$}
			\label{fig:centra}
		\end{figure}
		\begin{figure}[!htb]
			\centering
			\includegraphics[scale = 0.5]{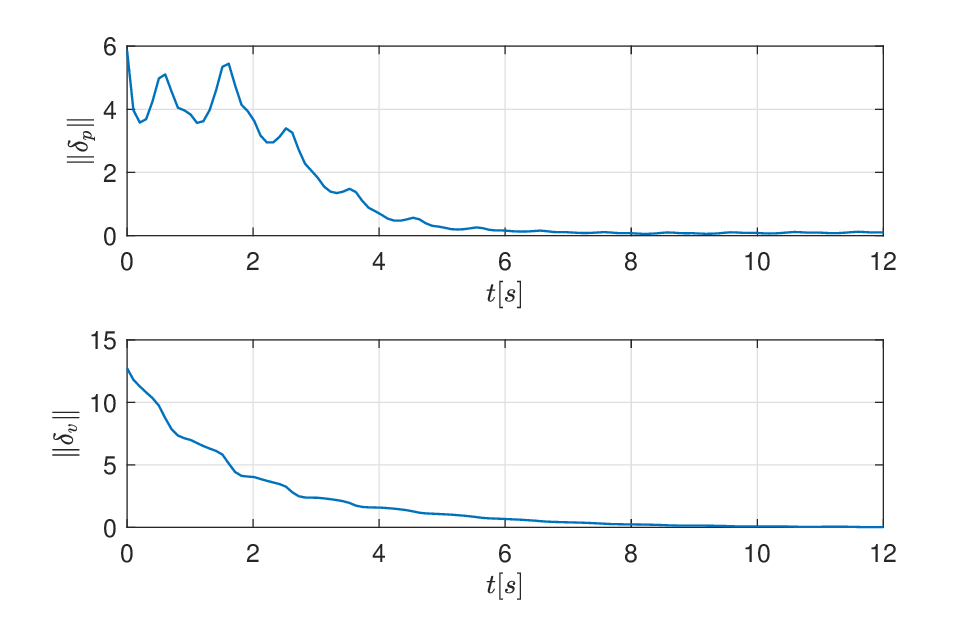}	
			\caption{Evolution of the estimation error $(\boldsymbol \delta_p,\boldsymbol \delta_v)$ under the decentralized observer with measurement noises. Gains are chosen as: $\kappa_k=10,\ M_k(0)=100I_{2d},\ Q_k=10I_{d}, \ S_k=0.01I_{2d}, \ \kappa_{o1}=10, \ \kappa_{o2}=5$.}
			\label{fig:decen}
		\end{figure}
		
		\begin{figure}[!htb]
			\centering
			\includegraphics[scale = 0.5]{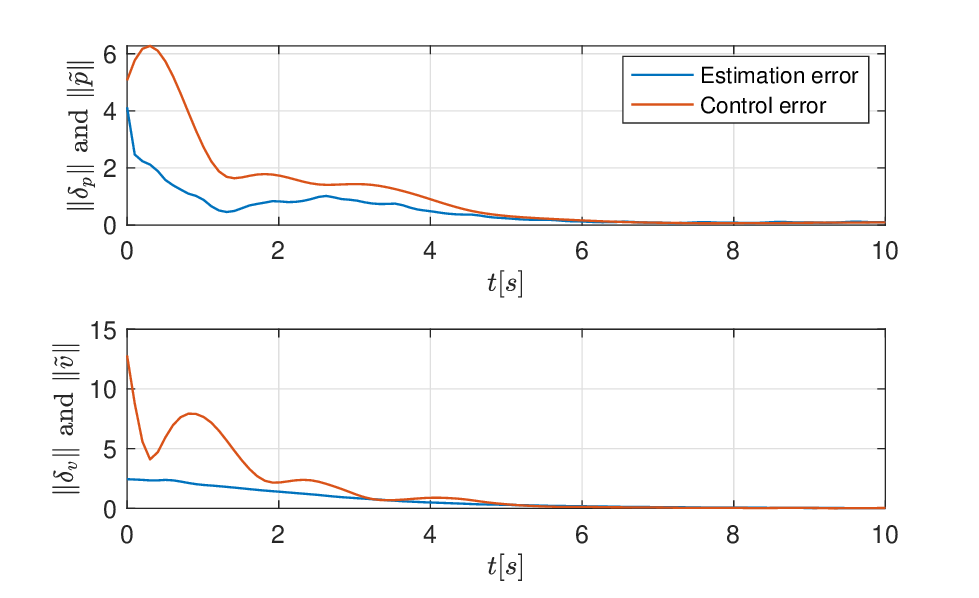}	
			\caption{Evolution of the estimation and control errors $(\boldsymbol \delta_p,\boldsymbol \delta_v)$, $(\tilde{\boldsymbol p},\tilde{\boldsymbol v})$. Gains for the decentralized observer are the same as in Fig. \ref{fig:decen}. Gains for controller: $k_{p_i}=5$ and $k_{p_i}=2$. }
			\label{fig:obs-control}
		\end{figure}
		This section provides simulation results to illustrate the effectiveness of the proposed observers and distributed observer-based formation tracking controller. All simulated bearing measurements $g_{ij}^m$ are deviated from their actual value $g_{ij}$ by random noise by defining $g_{ij}^m=\frac{(I_3+(0.02w)\times)g_{ij}}{|(I_3+(0.02w)\times)g_{ij}|}$, where $w\in \mathbb{R}^3$ is a random vector drawn from the standard normal distribution.
		
		We consider a 4-agent system under the undirected interaction topology with the incidence matrix \scalebox{0.7}{$H=\begin{bmatrix}1& -1& 0&0\\0& 1& -1&0\\0& 0& 1&-1\\1& 0& 0&-1 \end{bmatrix}$}.
		
		For both centralized and decentralized observers, the positions of agents are chosen as $p_1=[r+\frac r 2 \sin(t/f),r+\frac r 2 \sin(t/f),0],\ p_2=[0,r,0], \ p_3=[0,0,0],\ p_4=[r,0, 0]$ where $r=2\sqrt{2}, f=\frac 1 {2\pi}$. In this formation, the bearings $g_{14}$ and $g_{12}$ are PE and $g_{34}$ and $g_{23}$ are constant. Initial conditions are chosen the same for both observers: $\hat p_1(0)=[0,1,0], \ \hat p_2(0)=[2,0,1], \ \hat p_3(0)=[0,-1,1], \hat p_4(0)=[0,0,0]$, $\hat v_1(0)=[0,0,0], \ \hat v_2(0)=[1,0,0], \ \hat v_3(0)=[1,-1,0], \hat v_4(0)=[0,1,0]$. Fig. \ref{fig:centra} and \ref{fig:decen} show the evolution of the estimation error $(\boldsymbol \delta_p,\boldsymbol \delta_v)$ under the centralized and decentralized observers, respectively. 
		
		For the distributed observer-based controller, the desired formation is chosen as $p_1^*=[r+\frac r 2 \sin(t/f),r+\frac r 2 \sin(t/f),0],\ p_2^*=[0,r,0], \ p_3^*=[0,0,0],\ p_4^*=[r,0, 0]$. The initial position and velocity vectors are: $ p_1(0)=[1,0,0], \  p_2(0)=[-1,1,1], \  p_3(0)=[0,1,0],  p_4(0)=[0,0,0]$, $ v_1(0)=[0,0,1], \ v_2(0)=[1,-1,-1], \  v_3(0)=[1,0,1],  v_4(0)=[0,0,0]$. The initial conditions of the state estimation were chosen in the same way as the previous observers.  The effectiveness of the observer-based controller's performance is shown in Fig. \ref {fig:obs-control}.
		\section{Conclusion}
		This paper proposes an observer-based formation tracking control approach for multi-vehicle systems with second-order motion dynamics. It assumes that all vehicles can sense the bearings relative to neighboring vehicles and that only the leader can access its global position. By exploring Riccati gain and properties of BPE formation, both centralized and decentralized position and velocity observers are designed for the multi-vehicle system. Global exponential stability of the origin of the estimation error is guaranteed under both centralized and decentralized observers, provided the current formation is BPE. Finally, we propose an observer-based controller relying solely on vehicles' estimated positions and velocities and prove the local exponential stability of the distributed observer-based tracking controller, provided that the desired formation is BPE. Simulation results are presented to illustrate the performance of the proposed observers and the observer-based tracking controllers.
		\vspace{0.5cm}
		\appendix
		\begin{lem}[\cite{HamSam2017}] \label{extension}
			Consider the following generic linear time-varying (LTV) system
			\begin{equation}\label{ltv}
				\dot{x}=A(t)x+B(t)u, \quad y=C(t)x 
			\end{equation}
			If \begin{enumerate}
				\item $C(t)=\Pi(t) \bar{C}$ with $\bar{C}$ a constant matrix,
				\item $A$ is constant and such that the pair $(A,\bar{C})$ is Kalman observable,
				\item all eigenvalues of $A$ are real,
				\item $\Pi(t) \Pi^\top(t)$ is persistently exciting according to \eqref{eq:pe}
			\end{enumerate}
			then the system \eqref{ltv} is uniformly observable  
		\end{lem}

		\addtolength{\textheight}{-12cm}   
		


		%
		%
		%
		%
		%
		%
		%

		\bibliography{bibliography}
		\bibliographystyle{IEEEtran}

	\end{document}